\documentclass[%
 reprint,
 amsmath,amssymb,
 aps,
]{revtex4-2}

\usepackage{graphicx}
\usepackage{dcolumn}
\usepackage{bm}

\usepackage{color}

\newcommand{\be}{\begin{equation}}
\newcommand{\ee}{\end{equation}}

\begin{document}

\preprint{APS/123-QED}

\title{Laboratory rivers extremize friction and are cosmological analogues}

\author{Valerio Faraoni}
\email{vfaraoni@ubishops.ca}
\author{Nikki Veilleux}%
 \email{nveilleux21@ubishops.ca}
\affiliation{Department of Physics and Astronomy, Bishop's University, 2600 College Street, 
Sherbrooke, Quebec, Canada J1M~1Z7}


\date{\today}

\begin{abstract} 

In the shallow water approximation, the cross-sectional profiles of 
laboratory rivers satisfy a differential equation here shown to be 
formally the Friedmann equation of cosmology ruling the evolution of 
Anti-de Sitter universe. The ensuing cosmic analogy provides a 
counterintuitive Lagrangian for the transverse river profile. Extremizing 
the corresponding action corresponds to extremizing the friction force on 
the river bottom and the energy dissipation rate. Analysis of the second 
variation establishes that this extremum is a maximum.

\end{abstract}

\maketitle


\section{\label{sec:1} Introduction}


Curved spacetime phenomena such as Hawking radiation from black holes, 
particle creation in the early universe, superradiance, or backreaction 
are unaccessible in the laboratory with current and foreseeable 
technology. 
Analogue systems reproduce features of gravity in the laboratory and 
include fluids \cite{Unruh:1980cg,Unruh:1994je,Visser:1997ux, 
Fischer:2001jz, Visser:2004zs,Slatyer:2005ty,capillary:2008, 
Weinfurtner:2010nu,Torres:2016iee, Patrick:2018orp, Patrick:2019kis, 
Rousseaux:2020trm}, Bose-Einstein condensates and ultra-cold systems 
\cite{Volovik:1997mtb, Jacobson:1998he, Pashaev:1998sk, Volovik:1999mj, 
Volovik:2000ua, Barcelo:2000tg, Garay:2000jj,Barcelo:2003et, condensed2, 
Eckel:2017uqx, Prain:2010zq, CM4, Volovik:2002ci, Volovik:2003fe, 
Fedichev:2003id, Fedichev:2003bv, Fedichev:2003dj, Fischer:2004bf, 
Cha:2016esj, 
Braden:2019vsw,MunozdeNova:2018fxv,Yatsuta:2020uec,Tolosa-Simeon:2022umw, 
Rosu:2020tov}, and optical systems \cite{Schutzhold:2001fw, Unruh:2003ss, 
Davis:2003qfo, Schutzhold:2004tv, Prain:2019jqk, optical1}.

There is another, much less studied, class of analogies between cosmology 
and geological systems, including glaciers \cite{Chen:2014fqa, 
Faraoni:2016lhd,Faraoni:2020ips,JOG2020}, equilibrium beach profiles in 
oceanography \cite{Faraoni:2019wof,JOcean2020}, the freezing of lakes 
\cite{Faraoni:2020iel,Niehof:2026hyy} and other cooling phenomena 
\cite{Faraoni:2020ujg}, lava flows \cite{Faraoni:2023kzf}, and earthquakes 
\cite{Faraoni:2020kir} (similar analogies exist in mathematics as well 
\cite{Faraoni:2021tnx,Faraoni:2020swm}), see Ref.~\cite{mybook} for a 
comprehensive discussion. Some geological systems can be studied using 
laboratory 
analogues which, by extension, may become analogue universes. For example, 
lava flows are studied in the laboratory (e.g., \cite{LiuMei89}). Here we 
report a new cosmological analogy between laboratory rivers, the study of 
which has a long history, and cosmology. The analogy provides a 
ompletely counterintuitive Lagrangian describing the flow; analyzing its 
physical 
meaning in the shallow water approximation used in these experiments is 
quite instructive and leads to the result that laboratory rivers in 
equilibrium at the threshold of sediment transport extremize 
the friction force of the water on the bed.

Laboratory rivers are created by discharging water on top of inclined 
planes, which form a ``river'' bed composed of uniform, erodible, 
impermeable granular materials simulating the sediment in alluvial rivers. 
In the lab, the parameters of these artificial rivers can be carefully 
controlled and varied independently from each other. Contrary to real 
rivers, which are always turbulent, laboratory rivers exhibit laminar 
flow. The fact that the liquid discharged shapes its paths in a way 
similar to real streams and rivers is due to the universality of scaling 
laws obeyed by the water carving its path through an erodible bed, for 
example Lacey's law \cite{Lacey1930} relating the width $w_0$ of a river 
channel with the square root of its discharge $Q$.  (However, in laminar 
flow, the exponents appearing in these scaling laws differ from those 
observed in real rivers.)

Let us review the laboratory river results that are relevant for the 
cosmological analogy. Consider a river on an inclined plane of constant 
slope $S$ and introduce Cartesian axes $x$ and $y$ in this plane, with the 
$x$-direction aligned along the flow and the $y$-direction perpendicular 
to it, while the $z$-axis is vertical. Liquid (e.g., water mixed with 
glucose to increase viscosity \cite{PhysRevE.87.052204}) is released at 
the top with constant discharge $Q$ and the river adjusts its width 
$w_0$, its longitudinal slope $S$, and its cross-sectional profile to 
adapt to the discharge.  A cross-section of the river spans the interval $ 
0\leq y \leq w_0$, where the river is supposed to have constant width 
$w_0$. The water depth at $y$ is $D(y)$ and we refer all quantities to a 
unit length of river in the $x$-direction, along which the river is 
uniform. Let the density of the fluid be $\rho_f$, $g$ be the acceleration 
of gravity, and $\mu $ be the Coulomb friction coefficient for the 
granular material composing the bed.

There is a microscopic characteristic length scale in the river, of the 
order of the grain sediment size \cite{PhysRevE.87.052204}, 
\be
L = \frac{ \theta_t \left( \rho_s-\rho_f \right)d_s}{\mu \, \rho_f} \,,
\ee
where $\rho_s$ is the density of the sediment material, $d_s$ is the  
diameter of the spherical sediment grains, and $\theta_t$ is the 
dimensionless Shields  parameter appearing in the theory of sediment 
transport \cite{PhysRevE.87.052204}. The typical macroscopic scale of the 
river channel is $L/S $.

Assuming the shallow water approximation (which reads $|D'|\ll 1$ or 
$S\ll1$, making the channel scale $L/S \gg L$), the water depth $D(y)$ 
satisfies the ordinary differential equation \cite{PhysRevE.87.052204}
\be
\left( \frac{D'}{D}\right)^2 =\frac{ \mu^2}{D^2} -\alpha^2  
\label{depthODE}
\ee
where $\alpha \equiv S/L$, which admits the sine solution 
(Appendix~\ref{Appendix})
\be
D(y) = \frac{ \mu L}{S} \, \sin\left( \frac{Sy}{L} \right)  
\label{cos-solution}
\ee
satisfying $D(0)=D(w_0)=0$ provided that $\alpha w_0 =\pi $, which 
determines the width of the channel 
\be
w_0 = \frac{\pi L}{S}
\ee
in terms of the river parameters (Eq.~(13) of 
Ref.~\cite{PhysRevE.87.052204}).

Another (trivial) solution is the constant $D(y)=\mu/\alpha$, 
corresponding to a rectangular channel. It is also possible to join these 
two solutions creating a channel that is flat near its center and joined 
to two sine solutions (\cite{Henderson61}, such a solution is also 
mentioned in \cite{PhysRevE.87.052204}). However, here we want to describe 
laboratory rivers using a variational principle, which requires solutions  
to be continuous with their first {\em and second} derivatives (${\cal 
C}^2$), while such a composite solution would only be of class ${\cal 
C}^1$.

In the next section, we show that the shallow water 
equation~(\ref{depthODE}) is analogous to the Friedmann equation of 
general-relativistic cosmology describing the dynamical evolution of a 
universe under special conditions (negative spatial curvature and 
cosmological constant), and that the sine solution~(\ref{cos-solution}) 
corresponds to the Anti-de Sitter universe solving the corresponding 
Einstein-Friedmann equations (although this mathematical solution does not 
describe our observed universe, it is fundamental in string theories 
\cite{Hubeny:2014bla}).

\section{\label{sec:2} Cosmological analogy}

Let us introduce briefly the basics of general-relativistic 
cosmology \cite{footnote1}. The observed cosmic microwave background  
and large-scale structure surveys tell us that, on large scales, the 
universe is spatially homogeneous and isotropic, which translates in 
strong constraints on the solution of the Einstein equations of 
general relativity, which then is necessarily given by the 
four-dimensional  
Friedmann-Lema\^itre-Robertson-Walker (FLRW) line element  
\cite{Misner:1973prb,Wald,Carroll,KT,Liddle}  
\begin{eqnarray}
ds^2 &=& g_{\mu\nu} \, dx^{\mu} dx^{\nu} \nonumber\\
&&\nonumber\\
&=& -dt^2 +a^2(t) \left[ \frac{dr^2}{1-Kr^2} +r^2 \left( d\vartheta^2 + 
\sin^2 \vartheta \, d\varphi^2 \right)\right] \nonumber\\
&& \label{eq:10}
\end{eqnarray}
in comoving polar coordinates $x^{\mu}= \left(t, r, \vartheta, \varphi 
\right)$. The three-dimensional space can only have one of three 
geometries, corresponding to the sign of the constant $K$ (curvature 
index), which describes the curvature of the three-dimensional spatial 
sections obtained by setting $dt=0$. When $K>0$, the universe is closed;  
if $K=0$, 3-space is Euclidean; if $K<0$, 3-space has hyperbolic 
geometry \cite{Wald,Carroll,Liddle,KT}.

The expansion history of the cosmos is described by the scale factor 
$a(t)$. According to the Einstein equations, spacetime is curved by its 
mass-energy content. Different forms of matter produce different cosmic 
expansion histories $a(t)$.  In cosmology, the matter content of the 
universe usually consists of a perfect fluid with energy density $\rho(t)$ 
and isotropic pressure $P(t)$ related by a barotropic equation of state 
$P=P(\rho)$ (usually $P=w\rho$ with $w=$~const.).

Given the highly symmetric line element~(\ref{eq:10}), the Einstein 
equations reduce to the Einstein-Friedmann equations for the scale factor 
$a(t)$ and the matter variables $\rho(t), P(t)$ 
\cite{Wald,Carroll,Liddle,KT}
\begin{eqnarray}
&&H^2 \equiv \left( \frac{\dot{a}}{a}\right)^2 =\frac{8\pi G}{3} \, \rho 
-\frac{K}{a^2} \,, \label{eq:11}\\
&&\nonumber\\
&&\frac{\ddot{a}}{a}= -\, \frac{4\pi G}{3} \left( \rho +3P \right) \,, 
\label{eq:12} \\
&&\nonumber\\
&& \dot{\rho}+3H\left(P+\rho \right)=0 \,.\label{eq:13}
\end{eqnarray}
An overdot denotes differentiation with respect to the comoving time $t$, 
while $H(t)\equiv \dot{a}/a$ is the Hubble function 
\cite{Wald,Carroll,Liddle,KT}. These three equations are not all 
independent: given any two of them, the third one can be derived from 
them. In the following we choose the Friedmann equation~(\ref{eq:11}) and 
the energy conservation equation~(\ref{eq:13}) as primary equations, then 
the acceleration equation~(\ref{eq:12}) follows from them.  The analogy 
between laboratory rivers and FLRW cosmology holds if, in addition to the 
Friedmann equation~(\ref{eq:11}), the cosmic fluid obeys the energy 
conservation equation~(\ref{eq:13}), which is guaranteed if it has 
barotropic equation of state $P=w\rho$ with $w=$~const. Then, 
Eq.~(\ref{eq:13}) gives the scaling of the energy density
\be
\rho(a)=\frac{\rho^{(0)}}{  a^{3(w+1)}  } 
\,,\label{eq:14} 
\ee 
where the initial conditions determine the positive integration constant 
$\rho^{(0)} $ \cite{Misner:1973prb,Wald,Carroll,Liddle,KT}. Establishing 
an analogy 
is equivalent to establishing the validity of an equation of the 
form~(\ref{eq:11}), plus Eq.~(\ref{eq:14}) for the effective fluid source.
Einstein's famous cosmological constant $\Lambda$ can be incorporated into 
the Einstein-Friedmann equations by formally treating it as a fluid with 
energy density $\rho_{\Lambda} = \frac{\Lambda}{8\pi G}$ and  pressure 
$P_{\Lambda} =-\rho_{\Lambda}$.

It is well known in cosmology that the Einstein-Friedmann equations are 
obtained by varying the action of general relativity with a perfect fluid 
source \cite{Wald,Carroll}
\begin{eqnarray}
S_\mathrm{GR} &=&\int d^4 x \, \sqrt{-g^{(4)}} \left( \frac{R}{16\pi G} 
+\rho  \right) 
\nonumber\\
&&\nonumber\\
&=& 4\pi \int dr\, \frac{r^2}{\sqrt{1-Kr^2}}   \,\int dt \,  L\left( a, 
\dot{a} \right) \,, 
\end{eqnarray}
where $R$ is the Ricci scalar and $g^{(4)}$ is the determinant of the 
metric $g_{\mu\nu}$. The corresponding Euler-Lagrange (or the Hamilton) 
equations reproduce the Einstein-Friedmann equations. These Lagrangian and 
Hamiltonian are \cite{Faraoni:2004pi} 
\begin{eqnarray}
L \left(a, \dot{a} \right) &=& \frac{3}{8\pi G} \left( a \dot{a}^2 -Ka 
\right) +a^3 \rho \,,\\
&&\nonumber\\
{\cal H} \left(a, \dot{a} \right) &=& \frac{3}{8\pi G} \left( a \dot{a}^2 
+ Ka \right) -a^3 \rho \,,
\end{eqnarray}
while the dynamics is constrained and the ``scalar'' or ``Hamiltonian'' 
constraint ${\cal H}=0$ must be satisfied 
\cite{Misner:1973prb,Wald,Carroll}. The analogy between laminary 
laboratory rivers and FLRW cosmology can now be formulated.

The Friedmann equation in the absence of matter and with cosmological 
constant $\Lambda$ and spatial curvature,
\be
H^2=\frac{\Lambda}{3} -\frac{K}{a^2} \,,
\ee
is formally analogous to Eq.~(\ref{depthODE}) by setting 
\begin{eqnarray}
t \to y \,, & & a(t) \to D(y) \\
&&\nonumber\\
\Lambda &=& -3\alpha^2 <0 \,,\\
&&\nonumber\\
K &=& -\mu^2<0 \,,
\end{eqnarray}
where, in order to have real solutions, it must be $\lambda \geq 3K/a^2$ 
at all times $t$.
The Lagrangian for this system is 
\be
L\left( D, D' \right) = DD'^2 -\alpha^2 D^3 +\mu^2 D \,, 
\label{Lagrangian}
\ee
while the corresponding Hamiltonian is 
\be
{\cal H}\left( D, \pi_D\right)  = DD'^2 +\alpha^2 D^3 -\mu^2 D 
\,,\label{Hamiltonian} 
\ee
where $\pi_D \equiv \partial L/\partial D' = 2DD'  $ is the momentum 
canonically conjugated 
to $D$. Since the Hamiltonian~(\ref{Hamiltonian}) does not depend explicitly on 
$y$, it is conserved (Beltrami identity). Setting ${\cal H}=0$ yields  
the  equation for transverse laboratory river profiles 
\be
 D'^2 + \alpha^2 D^2 = \mu^2 \,.\label{questa}
\ee
Its solution~(\ref{cos-solution}) describes the Anti-de Sitter universe, 
which does not describe our universe but is of fundamental importance in  
string theories \cite{Hubeny:2014bla}. 

 Equation~(\ref{questa}) can be seen as the conservation 
equation 
$ T+V=\mu^2/2$ for a free, undamped harmonic oscillator consisting of a 
particle of unit mass at position $D(y)$ dependent on the ``time'' $y$, 
kinetic energy $T= D'^2/2$, potential energy $V=\alpha^2 D^2 /2$, and 
total mechanical energy $\mu^2/2$. The harmonic oscillator analogy is not 
useful while it is much more productive to discuss the cosmic analogy, 
which we do here.

The constant solution $D=\mu/\alpha$ corresponds to an unphysical static 
universe with constant $a=3K/\Lambda$, which is only realized by 
fine-tuning the cosmological parameters.

\section{\label{sec:3} Extremizing friction}

The question arises: what is, in the river context, the physical meaning 
of extremizing the action given by the gravitational analogy? In the 
following we show that the variational principle $\delta S_\mathrm{GR}=0$ 
corresponds 
to extremizing the friction force of the water on the river bed.  

The transverse profile of the river is described by the depth $D(y)$, 
the  line element along this  curve is 
\be
d\ell =\sqrt{dy^2+dD^2}=\sqrt{ 1+\left( 
\frac{dD}{dy} \right)^2}\, dy \equiv \sqrt{1+D'^2} \, dy
\ee
and, since we refer to a unit length of river in the $x$-direction, an elementary 
segment of the transverse river profile between $y$ and $y+dy$ spans the 
area parallel to the flow along the river bed $dA_{\parallel} 
=\sqrt{1+D'^2} \, dy$. According to the theory of Glover and 
Florey \cite{GloverFlorey1951}, at the threshold of sediment transport the 
stress at position $y$ along the river bottom is obtained by 
balancing gravity with the friction force, which yields (Eq.~(6) of 
\cite{PhysRevE.87.052204})
\be
\tau= \rho_f g  S D(y) \,.
\ee
The infinitesimal friction force on the area $dA_{\parallel} $ is then
\be
\tau dA_{\parallel} =\rho_f g S D(y) \sqrt{ 1+D'^2} \, dy\,.
\ee
The total friction force is obtained integrating along the river cross-section
\be
\int_0^{w_0} \tau dA_{\parallel} =\rho_f g S \int_0^{w_0} D\sqrt{1+D'^2} \, dy \,.
\ee 
Dropping irrelevant constants, the total friction force is extremized when 
the action integral 
\be 
J_0  \equiv \int_0^{w_0} D\sqrt{1+D'^2} \, dy
\ee
is extremized. This is the well-known action describing the classic 
brachistochrone variational problem, which has a hyperbolic cosine as its 
solution \cite{Goldstein,Boas} and its own soap film analogy 
\cite{CriadoAlamo}.  (The slightly more general Lagrangian $\bar{L}= D^k 
\sqrt{1+D'^2}$ appears (without Lagrangian constraints) in a variational 
principle describing equilibrium beach profiles 
\cite{Larson99,JenkinsInman06, Maldonado20, 
MaldonadoUchasara06,Faraoni:2019wof}. In this context, $D(y)$ is the water 
depth as a function of the perpendicular distance $y$ to the shore, and 
the energy dissipation rate due to waves on the sea bottom is minimized 
\cite{JOcean2020}. However, $J_0$ cannot be the full action integral 
because, to be true to the physical problem at hand, one must impose 
conservation of mass (or, since the fluid is assumed to be incompressible, 
of volume) and constant discharge $Q$, which is obtained by imposing two 
Lagrangian constraints.

To enforce  mass conservation, consider again a unit of river length in 
the $x$-direction and the plane $\left(y, D(y) \right)$ transverse to the 
flow. The transverse area occupied by the fluid is
\be
A_{\perp} = w_0 D_\mathrm{max} -\int_0^{w_0} D(y) dy \,,
\ee
where $D_\mathrm{max}$ ({\em a posteriori} equal to  $\mu L/S $) is the 
maximum depth. Therefore, we add to the action $J_0$ the integral
$ J_1 \equiv -\lambda_1 \int_0^{w_0} D(y) dy$, where $\lambda_1$ is a 
dimensionless  Lagrange multiplier. 

The river discharge (precisely controlled in laboratory experiments) is 
\be
Q= \int_0^{w_0} U(y) D(y) dy \,,
\ee
where $U(y)$ is the vertically averaged fluid velocity at $y$. At 
the 
threshold of sediment transport in the laminar regime considered, it is  
(Eq.~(10) of Ref.~\cite{PhysRevE.87.052204}) 
\be
U(y) =\frac{g S}{3\nu} \, D^2(y) \,,
\ee
where $\nu$ is the kinematic viscosity coefficient and 
\be
Q= \int_0^{w_0} U(y) D(y) dy = 
\frac{g S}{3\nu} \,  \int_0^{w_0} D^3(y) dy \,.
\ee
The constancy of this discharge is enforced by adding the term $J_2 \equiv 
-\lambda_2  
  \int_0^{w_0} D^3(y) dy $ to the action, where $\lambda_2$ is a second 
Lagrange multiplier with the dimensions of the inverse of an area.  To 
conclude, the total friction force is extremized subject to conservation 
of mass and discharge by imposing the variational principle $\delta J =0$ 
(with vanishing variations at the endpoints $y=0, w_0$) on the action 
integral
\begin{eqnarray}
J\left[ D(y) \right] &=&  J_0 + J_1 + J_2 \nonumber\\
&&\nonumber\\
&=& \int_0^{w_0}  \left[ D\sqrt{1+D'^2} -\lambda_1 D -\lambda_2 D^3 \right] 
dy\nonumber\\
&&\nonumber\\
& \equiv & \int_0^{w_0} L\left( D, D' \right) dy  \,,
\end{eqnarray}
a functional of the transverse river profile $D(y)$. The Euler-Lagrange 
equation 
\be
 \frac{d}{dy} \left( \frac{ \partial L}{\partial D'} \right) - \frac{\partial 
L}{\partial D} =0
\ee
for the Lagrangian
\be
L\left( D, D' \right) = D\sqrt{1+D'^2} -\lambda_1 D -\lambda_2 D^3 
\ee
are rather complicated but the discussion of laboratory rivers uses the 
shallow water approximation \cite{PhysRevE.87.052204} corresponding to 
$|D'|\ll 1$. In this approximation, the square root can be expanded giving 
the approximate Lagrangian (apart from an irrelevant overall factor~2)
\be
L \simeq DD'^2 -\alpha^2 D^3 +\mu^2 D \,,\label{approxLagrangian}
\ee
where
\begin{eqnarray}
\alpha^2 &=& 2\lambda_2 \,,\\
&&\nonumber\\
\mu^2 &=& 2\left( 1- \lambda_1 \right) \,,
\end{eqnarray}
which implies that $ \lambda_1<1$ and $\lambda_2>0$ and provides the 
physical interpretation of the Lagrange multipliers $\lambda_{1,2}$. The 
Lagrangian~(\ref{approxLagrangian}) is nothing but the 
Lagrangian~(\ref{Lagrangian}) obtained by the analogy between laminar 
laboratory rivers and cosmology. Therefore, laboratory rivers in 
equilibrium extremize the friction force of the water on the river bed.

\section{\label{sec:4} Maximum or minimum?}

The solution~(\ref{cos-solution}) extremizes the action $J$. Variations 
$\eta(y)$ 
around the actual path are required to be continuous with their first and 
second derivatives on $\left[0, w_0\right]$ and to vanish at the 
endpoints,  $\eta(0)=\eta(w_0)=0$. They are parametrized by a parameter 
$\epsilon$ so that the varied  paths are  
\cite{WeberArfken}
\be
\bar{D}\left( y, \epsilon \right)= D(y)+\epsilon \, \eta(y) \,.
\ee
Then we have 
\begin{eqnarray}
\bar{D}'\left( y, \epsilon \right) &=& D'(y)+\epsilon  \, 
\frac{d\eta}{dy} \,,\\
\frac{\partial \bar{D}}{\partial  \epsilon } &=& \eta \,,\quad 
\frac{\partial \bar{D}'}{\partial \epsilon } = 
\frac{d\eta}{dy} \,,\\
\frac{\partial^2 \bar{D}'}{\partial \epsilon^2}&=& 
\frac{\partial^2 \bar{D}}{\partial \epsilon^2}=0 \,,
\end{eqnarray}
and one obtains \cite{WeberArfken}
\be
\frac{\partial^2 J}{\partial \epsilon ^2} = \int_{0}^{w_0}dy \left[
\frac{\partial^2 L}{\partial D'^2} \left( \frac{ d\eta}{dy}\right)^2 
+  \frac{\partial^2 L}{\partial D \partial D'} \, \frac{ 
d(\eta^2)}{dy} + \frac{\partial^2 L}{\partial D^2} \, \eta^2 \right] \,. 
\label{integral}
\ee
Inserting the derivatives 
\begin{eqnarray}
\frac{\partial^2 L}{\partial D^2} & = & -6\alpha^2 D  \,,\\
&&\nonumber\\
\frac{\partial^2 L}{\partial D\partial D'}&=& 2D'   \,,\\
&&\nonumber\\
\frac{\partial^2 L}{\partial D'^2}&=&  2D  \,,
\end{eqnarray}
into~(\ref{integral}) gives
\be
\frac{\partial^2 J}{\partial \epsilon ^2} = 2\int_{0}^{w_0}dy \left[
D \left( \frac{ d\eta}{dy}\right)^2 
+  D' \, \frac{ d(\eta^2)}{dy} -3\alpha^2  D \, \eta^2 \right] \,. 
\label{integral2}
\ee
To compute this integral 
along the path that extremizes $J$, we use the equation  
$D''=-\alpha^2 D $ satisfied by this path (which is obtained by 
differentiating Eq.~(\ref{questa})). Then, we integrate by parts the term
\begin{eqnarray}
\int_0^{w_0} dy \, D' \, \frac{d(\eta^2)}{dy} &=&\Bigg[ \eta^2 D' 
\Bigg]_0^{w_0} - \int_0^{w_0} dy \,\eta^2  D'' \nonumber\\
&&\nonumber\\
&=& \alpha^2 \int_0^{w_0} dy \,  D \eta^2  \,,
\end{eqnarray}
using the fact that $\eta(y)$ vanishes at the endpoints.
Putting everything together, we have
\begin{eqnarray}
\frac{\partial^2 J}{\partial \epsilon^2} = 2\int_0^{w_0} dy \, D(y) 
\left[
\left( \frac{d\eta}{dy}\right)^2 -2\alpha^2 \eta^2 \right]  \,.
\end{eqnarray}
Unfortunately, the two terms of opposite signs in the square 
brackets of the integrand have 
similar magnitudes and none of them dominates, which makes it difficult 
to assess the sign of $\partial^2 J/\partial \epsilon^2$. However, since 
the action is extremized for all varied paths that are ${\cal C}^2$ and 
vanish at the endpoints, we can determine the sign of $\partial^2 
J/\partial \epsilon^2$ on a particular varied path. We choose
\be
\eta_1(y) = 1-\cos(2\alpha y) \,.
\ee
Then 
\begin{eqnarray}
\frac{\partial^2 J}{\partial \epsilon^2} &=&  2\int_0^{w_0} dy \, 
\frac{\mu}{\alpha} \, \sin(\alpha y) \left\{ 
4\alpha^2 \sin^2(2\alpha y) \right.\nonumber\\
&&\nonumber\\
&\,& \left. -2\alpha^2 \left[ 1+\cos^2(2\alpha y) 
-2\cos(2\alpha y) \right] \right\} \nonumber\\
&&\nonumber\\
&=& -\frac{128\mu}{15}\,,
\end{eqnarray}
where we used the fact that $\alpha w_0=\pi$. The second variation 
$ \frac{\partial^2 J}{\partial \epsilon^2}\Bigg|_{\epsilon=0}  $ is 
negative, therefore the 
path $D(y)$ extremizing $J$ corresponds to a {\em maximum} of the action  
$J$ and to maximum friction force along the river bed. 

By choosing the other variation 
\be
\eta_2(y) = A y \left( y- \frac{\pi}{\alpha} \right) 
\ee
(where $A$ is a constant), one easily computes 
\begin{eqnarray}
\frac{\partial^2 J}{\partial \epsilon^2} &=& \frac{2 \mu A^2}{\alpha}
\int_0^{w_0} dy \, \sin(\alpha y) \Bigg[
-2\alpha^2 y^2 +4\pi \alpha y^3 \nonumber\\
&&\nonumber\\
& \, &  +\left( 4-2\pi^2\right) y^2 
-\frac{4\pi}{\alpha} \, y +\frac{\pi^2}{\alpha^2} \Bigg] \nonumber\\
&&\nonumber\\
&=& \frac{4 \mu A^2}{\alpha}\left( 5\pi^2 -56\right) \,,
\end{eqnarray}
which is again negative, validating the maximum of $J$.

\section{\label{sec:5} Discussion and conclusions}

The shallow water approximation and the assumption that the river is in 
equilibrium at the threshold of sediment transport are not satisfactory 
for real alluvial rivers, which usually actively transport sediment. 
Moreover, the discharge $Q$ and the slope $S$ are usually variable, and 
the sediment is heterogeneous. Nevertheless, by isolating and controlling 
parameters and idealizing the possible physical situations, laboratory 
rivers give us an opportunity to understand the basic mechanisms ruling 
rivers.

Laminary laboratory rivers in equilibrium at the threshold of sediment 
transport are analogous to a particular universe which is empty of matter, 
is dominated by a negative cosmological constant, and is negatively 
curved, i.e., the famous Anti-de Sitter universe of string theories and 
the 
AdS/CFT correspondence. The cosmological analogy suggests an effective, 
and counterintuitive, Lagrangian for the differential 
equation~(\ref{depthODE}) obeyed by the transverse river profile. The 
question of what is the physical meaning of the action principle 
associated with this Lagrangian turns out to be useful to understand the 
physics of laboratory rivers: it corresponds to extremizing the friction 
force of the water against the river bottom, subject to the constraints of 
mass conservation and constant discharge, which are enforced by Lagrange 
multipliers. Since the rate of energy dissipation due to the friction 
force 
$\vec{F}$ along the river bed is $ dE/dt= \vec{F}\cdot \vec{v}=Fv$ and the 
velocity of the water is constant, extremizing the friction force 
corresponds to extremizing the energy dissipation rate (since the river is 
uniform along the $x$-direction, its kinetic energy density does not 
change and heat is generated only at the expense of gravitational 
potential energy). The question arises of whether the extremized 
dissipative force (or the energy dissipation rate)  is maximum or minimum. 
This question is settled by examining the second variation $\delta^2 J$, 
establishing that the total friction force on the river bed and the energy 
dissipation rate are {\it maximized}.


\begin{acknowledgments}
 
V.F. is partially supported by the Natural Sciences \& Engineering 
Research Council of Canada (Grant No.~2023-03234).

\end{acknowledgments}


\appendix
\section{\label{Appendix} Proof of Eq.~(\ref{cos-solution})}

Equation~(\ref{depthODE}), rewritten as
\be
D'^2=\mu^2-\alpha^2 D^2 \,,
\ee
gives
\be
\frac{D'}{ \sqrt{ \left( \mu/\alpha \right)^2 -D^2}} = \pm \alpha
\ee
and
\be
\int \frac{dD}{ \sqrt{ \left(\mu/\alpha \right)^2-D^2}} = 
\pm \alpha \left(y-y_0\right) \,,
\ee
which integrates to
\be 
\arctan \left( \frac{D}{ \sqrt{ \left(\mu/\alpha \right)^2-D^2}} \right) = 
\pm \alpha \left(y-y_0\right) \,.\label{integrated}
\ee
Imposing $D(0)=0$ gives $y_0=0$, while the second boundary condition 
$D(w_0)=0$ yields 
\be
w_0= \frac{\pi}{\alpha} =\frac{ \pi L}{S} \,.
\ee
Taking the tangent of both sides of Eq.~(\ref{integrated}) and  
squaring gives
\be
D^2= \left( \frac{\mu^2}{\alpha^2} -D^2 \right) \tan^2 (\alpha y) \,;
\ee
collecting the two terms in $D^2$,
\be
D^2 =\frac{\mu^2}{\alpha^2} \, \sin^2 (\alpha y) \,,
\ee
from which Eq.~(\ref{cos-solution}) follows.




\begin{thebibliography}{99}
\bibitem{footnote1} We adopt the notation of Ref.~\cite{Wald}, using units 
in which the speed of light $c$ is unity. The metric tensor $g_{\mu\nu}$ 
has signature ${-}{+}{+}{+}$ and $G$ is Newton's constant.
\end{thebibliography}

\end{document}